# Controlling transport properties of graphene nanoribbons by codoping-induced edge distortions

Hyo Seok Kim, Seong Sik Kim, Han Seul Kim, and Yong-Hoon Kim*

Graduate School of Energy, Environment, Water, and Sustainability, Korea Advanced Institute of Science and Technology, 291 Daehak-ro, Yuseong-gu, Daejeon 305-701, Korea.

One notable manifestation of the peculiar edge-localized states in zigzag graphene nanoribbons (zGNRs) is the *p*-type (*n*-type) characteristics of nitrogen (boron) edge-doped GNRs, and such behavior was so far considered to be exclusive for zGNRs. Carrying out first-principles electronic structure and quantum transport calculations, we herein show that the donor-acceptor transition behavior can also arise in the B/N edge-doped *armchair* GNRs (aGNRs) by introducing a bipolar P codopant atom into the energetically most favorable nearest neighbor edge sites. The *n*-type (*p*-type) transport properties of B,P (N,P) co-doped aGNRs are also shown to be superior to those of reference single N (B) doped aGNRs in that the valence (conduction) band edge conductance spectra are better preserved. Disentangling the chemical doping and structural distortion effects, we will demonstrate that the latter plays an important role in determining the transport type and explains the donor-acceptor transition feature as well as the bipolar character of P-doped aGNRs. We thus propose the systematic modification of GNR edge atomic structures via co-doping as a novel approach to control charge transport characteristics of aGNRs.

## Introduction

A promising direction to realize advanced graphene-based electronic devices [1-3] is to open a band gap by forming graphene nanoribbons [4,5] and tailoring their properties via doping [1-3]. In terms of the synthesis of GNRs, significant progress was recently made in both top-down [4,5] and bottom-up approaches [6-15]. The latter has particularly led to atomically precise narrow armchair GNRs (aGNRs), providing valuable information on the intrinsic properties of GNRs [6-12] and even the possibility of making reliable junction structures [13-15]. The next natural experimental target would be to realize atomically controlled doped GNRs, and recently N-doped aGNRs [12,14,16,17] and B-doped aGNRs [16-17] have been indeed successfully synthesized. In this context, we believe that it is desirable to revise and expand earlier mostly theoretical studies on the doping of GNRs [18-25].

Herein, based on combined density functional theory (DFT) and DFT-based non-equilibrium Green's function (NEGF) calculations, we study the structural, electronic, and charge transport properties of binary B,P and N,P edge-co-doped aGNRs. Previous theoretical research mostly focused on the GNRs doped by single elements [18-24], and multi-doping cases have been rarely considered [23,25]. In this paper, we start from the representative *p*-type (acceptor character) B-doped and *n*-type (donor character) N-doped aGNRs and study the B,P and N,P co-doping effects. The P-doped aGNRs are known to exhibit a bipolar character, and moreover they involve strong $sp^3$ structural distortions [26,27]. Surprisingly, we find that the P co-doping converts the *p*-type (*n*-type) B-doped (N-doped) aGNRs into *n*-type (*p*-type). This anomalous donor-acceptor transition behavior was previously observed only for zigzag GNRs (zGNRs) that have edge localized states [18-24], and to our knowledge has not been reported for aGNRs. We will systematically analyze this phenomenon by decomposing the contributions from chemical doping and structural relaxation, and show that the latter plays a critical role in determining the transport characteristics of aGNRs. Moreover, in comparing the *n*-type (*p*-type) property of N (B) doping with that of B,P (N,P) doping, we will also point out hitherto undiscussed undesirable aspects of single N and B doping and argue that the B,P and N,P co-doping approach provides a more robust doping method. Most importantly, this work suggests controlled distortions of edge atomic structures of GNRs as a novel scheme to regulate their charge transport properties.

## Results

We first discuss the energetics of the B,P- and N,P-co-doped aGNRs. For the 11-aGNRs (width $W = 14.21$ Å) with various doping configurations (Figures 1a and 1b), we have calculated the relative formation energies,

$$E_{rf} = E_{doped} - E_{gs}, \qquad (1)$$

* To whom correspondence should be addressed. E-mail: y.h.kim@kaist.ac.kr

where $E_{doped}$ is the total energy of the aGNR in a given binary-doping configuration, and $E_{gs}$ is the total energy of the ground (lowest-energy) state binary-doped aGNR. The results are presented in Figure 1c: In line with the single B-, N-, and P-doped aGNR cases [24], we found that the aGNR edge sides are preferred doping sites for the binary B,P and N,P co-doping. Particularly, as in the binary B,N-doped aGNR case we have previously considered [25], we concluded that the most favorable binary doping configurations are obtained when the B (N) and P dopant atoms are bonded to each other and form a B-P (N-P) complex at the aGNR edges.

Figure 1b shows the fully optimized atomic structures of binary B,P- and N,P-doped 11-aGNRs, together with those from single B-, N-, and P-doped counterparts. We note that, unlike in the single B-, single N-, and binary B,N-doping cases [25], significant structural distortions appear in the single P-, binary B,P-, and binary N,P-doping cases. This effect, which also occurs in the substitutional single P doping and P-based co-doping of graphene basal plane and CNT [26,27], is due to the bigger atomic size of P and its $sp^3$ bonding character. Bond lengths involving P atoms are noticeably larger than the C-C $sp^2$ bond length of 1.42 Å: The C-P bond length ranged from 1.79 Å (P- and B,P-doping cases) to 1.81 Å (N-P doping case), and the B-P and N-P bond lengths were 1.82 Å and 1.70 Å, respectively. The C-P-C (P doping case), B-P-C (B,P doping case), and N-P-C (N,P doping case) bond angles were 99.7°, 102.9°, and 95.8°, respectively. Finally, we note the U- and Z-shape distortions of the B,P and N,P co-doped GNR edge atomic structures (Figure 1b top panels), which can be understood by the chemical valency of the B, N, and P atoms.

Because the structural distortion of aGNR edges caused by the co-doping involving a P atom will be shown to play a critical role in determining their transport characteristics, we have carefully confirmed the robustness of structural features (including the U- and Z-shape edge distortions in the B,P and N,P co-doped GNRs, respectively) and their effects on aGNR transport properties. We first reproduced the optimized geometries and transport data using the generalized gradient approximation [28] and our in-house code [29-31] (see Figure S3). Next, we also confirmed that structures and transmission functions are unaffected by adopting ten unit cells rather than six unit cells (see Figure S4). For the lowest energy (connected B-P and N-P) conformations identified above, we also carried out cell optimizations and found that optimal lattice parameters increase from the pristine GNR ones by the small amount of 0.2 Å and 0.1 Å for the B,P and N,P co-doping cases, respectively (see Figure 1d). This again only negligibly modified transmission functions (see Figures S5b and S5c). In Table I, we report the formation energies of B,P and N,P co-doped six-unit-cell 11-aGNRs calculated at the optimal lattice parameters according to

$$E_f = (E_{doped} + 2\mu_C) - (E_{pristine} + \mu_{B/N} + \mu_P), \quad (2)$$

where $E_{pristine}$ is the total energy of the pristine aGNR. The chemical potentials of C, B, N, and P, $\mu_C, \mu_B, \mu_N, \mu_P$, were extracted from the graphene, α-rhombohedral boron, $N_2$ molecule, black phosphorus, respectively. The results show that the N,P co-doping is an exothermic process, and, although it is an endothermic one, the B,P co-doping is energetically more favorable than the single P doping.

We next discuss the electronic and charge transport properties of the B,P-complex- and N,P-complex-doped aGNRs in comparison with the single B-, single N-, and single P-doped counterparts. As references, we first discuss the electronic band structures and corresponding transmission spectra of B-, N-, and P-edge-doped 11-aGNRs (Figures 2a-2c). As indicated in the fat band diagrams, defect states originating from the B/ N/ P dopant atom and the corresponding suppression of transmission spectrum from the pristine aGNR value of 1 $G_0$ (= $e^2$/h) appear below/above/around the Fermi level $E_F$, or the B, N, and P doping of aGNR edges induces the $p$-type, $n$-type, and bipolar characteristics, respectively [22,24,25].

At this point, we emphasize one undesirable aspect of the B and N doping of aGNRs, which could have important implications in the future GNR-based device applications but has not been explicitly discussed in the literature. Specifically, the single B (N) edge doping of aGNRs destroys the conductance (valence) band edge transmission to a high degree as well as the valence (conduction) band transmission, or the B (N) doping of aGNRs has a partial $n$-type ($p$-type) character as well as the main $p$-type ($n$-type) one. This feature results from the spatially and energetically delocalized nature of the B (N) defect states, which interferes with the major transport channel along the central region of aGNRs. As shown in Figure 3a for the N-doped aGNRs, these two effects become more pronounced in the narrower aGNRs (see Figure S6 for the complete survey of the N-doping aGNR case). Finally, we note that, while the bipolar character of the P-doped aGNR was reported earlier [24], its origin was not clearly explained.

Having pointed out several important aspects of the B, N, and P doping of aGNRs, we next move on to the B,P-

and N-P-complex edge-doped aGNR cases. The band structures and transmission spectra shown in Figures 2d and 2e demonstrate the following:

(1) Introduction of a bipolar P co-dopant atom next to the B (N) atom converts the B (N) edge-doped aGNR from $p$-type ($n$-type) to $n$-type ($p$-type). The energetic shift of the B/N-originated states can be clearly identified in the fat band diagrams (Figure 2a vs Figure 2d and Figure 2b vs Figure 2e).

(2) Compared with the N-doped (B-doped) aGNR counterpart, the valence (conduction) band edge conductance spectra are better preserved in the B,P-doped (N,P-doped) case. Namely, the $n$-type ($p$-type) character of the B,P (N,P) co-doped aGNRs is more reliable compared with that of the N-doped (B-doped) aGNRs.

As shown in Figures 3b and 3c, both of these effects are universal throughout all three aGNR families and robustly observed even for narrow-width aGNRs. For example, whereas the collapse of the valence band edge conductance is significant in the N-doped 9-aGNR, it is maintained close to 1 $G_0$ in the B,P-doped 9-aGNR.

## Discussion

We now analyze in detail the origin of anomalous doping effects observed in the binary B,P- and N,P-doped aGNRs. Particularly, noting that incorporating the P dopant atom at the GNR edge induces strong structural distortions in addition to the change in chemical valency, it would be helpful to isolate the contribution of structural distortion from the purely electronic structural change. For this purpose, we replaced the edge C atom by a P atom (P-doping) or the C atom next to B or N by a P atom (B,P- and N,P-doping, respectively) without allowing further structural relaxations, and next compared the results from these constrained "ideal" models with the corresponding data from fully-relaxed models (Figure 2). The band structures and charge transport characteristics of the constrained P-, B,P-, and N,P-doped aGNRs are shown in Figure 4. Overall, we find that the structural distortion of aGNR edges plays a critical role in determining the final "doping" effects: First, the constrained P-doped aGNR shows an $n$-type behavior as can be expected solely from the chemical valency of the P atom (Figure 4a). So, the structural distortion can be identified as the origin of the bipolar character (Figure 2c). Next, as in the B,N-doping case [25], the constrained B,P doping model shows an almost perfect recovery of the intrinsic graphene conductance (Figure 4b). This can be again expected from the valency of the connected B-P atoms, and we can conclude that the structural distortion destroys the conductance and induces the $n$-type behavior (Figure 2d). Finally, for the N,P doping case, we find a strong destruction of conductance throughout wide energy range around (Figure 4c). Namely, in this case, the structural distortion in fact restores the conductance throughout a wide energy range around $E_F$ (Figure 2e). So, we can conclude that the anomalous donor-acceptor transition reported in this work mainly originates from the co-doping induced structural distortions in the B,P doping case and combinatorial effect of electronic doping and structural distortion in the N,P doping case. We emphasize again that, once the co-doping pairs (B,P and N,P) are specified, the geometries of the distorted aGNR edges (and thus the doping effects) are robustly reproduced.

To further characterize the nature of B-P and N-P defect states identified in the relaxed and constrained 11-aGNR models, we present in Figure 5 their local density of states (LDOS) corresponding to the transmission dips in Figures 2 and 4 (shaded area). For the constrained B,P-doped aGNR model that exhibits an almost perfect transmission spectrum (Figure 4b), we showed the LDOS for the same energy region as the constrained N,P co-doping case (Figure 4c). Overall, we observe that the LDOS from the fully relaxed models exhibit less pristine-like LDOS. Namely, in the B,P and N,P co-doped aGNRs, structural distortions result in the more effective elimination of current-carrying states at the transmission dips, which on the other hand allows the better preservation of transmission spectra in neighboring energy states. We also note such an effective reduction of LDOS cannot be identified for the single B-, N-, and P-doping cases (see Figure S7 as well as Figure 2 of Ref. 25), which further clarifies the above-explained more robust doping effects achieved with the B,P and N,P co-doping approach.

## Conclusions

In summary, we reported for the first time that the $n$-type N ($p$-type B) edge-doped aGNR can be converted into $p$-type ($n$-type) upon introducing a bipolar P co-dopant atom and forming the N-P (B-P) complex. Formerly, such anomalous acceptor-donor transition was predicted only for zGNRs, and the Coulomb repulsion (attraction) between GNR zigzag edge localized states and B (N) dopant states was identified as its origin. In our aGNR cases, separating the structural distortion effects from the purely electronic ones, we showed that the robust $sp^3$-type distortions of GNR edge atomic structures play a critical role in determining the transport properties of B,P and N,P co-doped aGNRs.

With the proposed novel co-doping methods that involve the "structural doping" effect and particularly the experimental availability of atomically precise aGNRs [6-15] in mind, we have also revisited the issue of *n*-type (*p*-type) doping of aGNRs based on the single N (B) doping. Specifically, we argued that the aGNR co-doping approach introduced in this work achieves more precise doping effects by better preserving the valence and conduction band edge transmission spectra in the *n*- and *p*-type doping cases, respectively. Analyzing the mechanisms, we showed that the distortions in GNR edge atomic structures induced by the formation of B-P- and N-P complexes result in defect states that are spatially and energetically more localized than those of N and P that only involve purely electronic doping effects. In closing, we note that an experimental group has recently demonstrated the feasibility of generating systematically varying degree of GNR edge distortions by introducing different edge-doping elements [32].

## Methods

***DFT calculations.*** Labeling aGNRs by the number of dimer lines contained in their unit cell *N*, we considered hydrogenated $N = 9$, 10, and 11-aGNRs to cover all three families of aGNRs with band gaps in the order of $E_g^{3p+1} > E_g^{3p} > E_g^{3p+2}$ (*p* is an integer). The ground-state atomic and electronic structures of single B-, N-, and P- and binary B,P- and N,P-doped aGNRs were obtained through DFT calculations within the local density approximation [33] using the SIESTA software [34]. We adopted rectangular simulation boxes containing six unit-cell aGNR models (length $L = 25.56$ Å) with periodic boundary condition along the aGNR axial *z*-axis direction. A vacuum space of minimum 20.0 Å were inserted along the transverse *x*- and *y*-directions. The atomic cores were replaced by norm-conserving nonlocal pseudopotentials of Troullier-Martins type [35]. The double ζ-plus-polarization-level numerical atomic orbital basis sets defined by the 80 meV energy cut-off and the real-space mesh defined by 200 Ry kinetic energy cutoff were adopted.

***NEGF calculations.*** For the calculation of charge transport properties, after adding four additional pristine aGNR unit cells as electrodes (see Figure S1), we carried out fully self-consistent NEGF calculations using the TranSIESTA code [36]. The surface Green's functions were extracted from two independent DFT calculations for the two-unit-cell-long pristine aGNRs corresponding to electrodes 1 and 2 with a 96 $\vec{k}_\perp$-point sampling along the charge transport z direction (in Figure 1a). In calculating the transmission function T(E), the energy window was scanned from −1.5 to +1.5 eV near Fermi level at the 0.001 eV resolution.

## Acknowledgements

This research was supported mainly by the Global Frontier Hybrid Interface Materials Program (2013M3A6B1078881) of the National Research Foundation (NRF) funded by the Ministry of Science, ICT and Future Planning of Korea, and additionally by the NRF Nano·Material Technology Development Program (2012M3A7B4049888) and the KIST Institutional Program (2Z04490-15-112). Computational resources were provided by the KISTI Supercomputing Center (KSC-2014-C3-021), and SIESTA calculations were performed using the LCAODFTLab interface available at the EDISON Nanophysics site (https://nano.edison.re.kr).

## Author Information

**Contributions**
Y.-H.K. conceived and designed the project. Hyo S.K. and S.S.K. performed modeling and calculations, and prepared figures. Han S.K. helped in analyzing the data. Y.-H.K. wrote the manuscript with assistance by Hyo S.K. and S.S.K.

**Competing interests**
The authors declare no competing financial interests.

**Corresponding Author**
Correspondence to Yong-Hoon Kim. *E-mail: y.h.kim@kaist.ac.kr

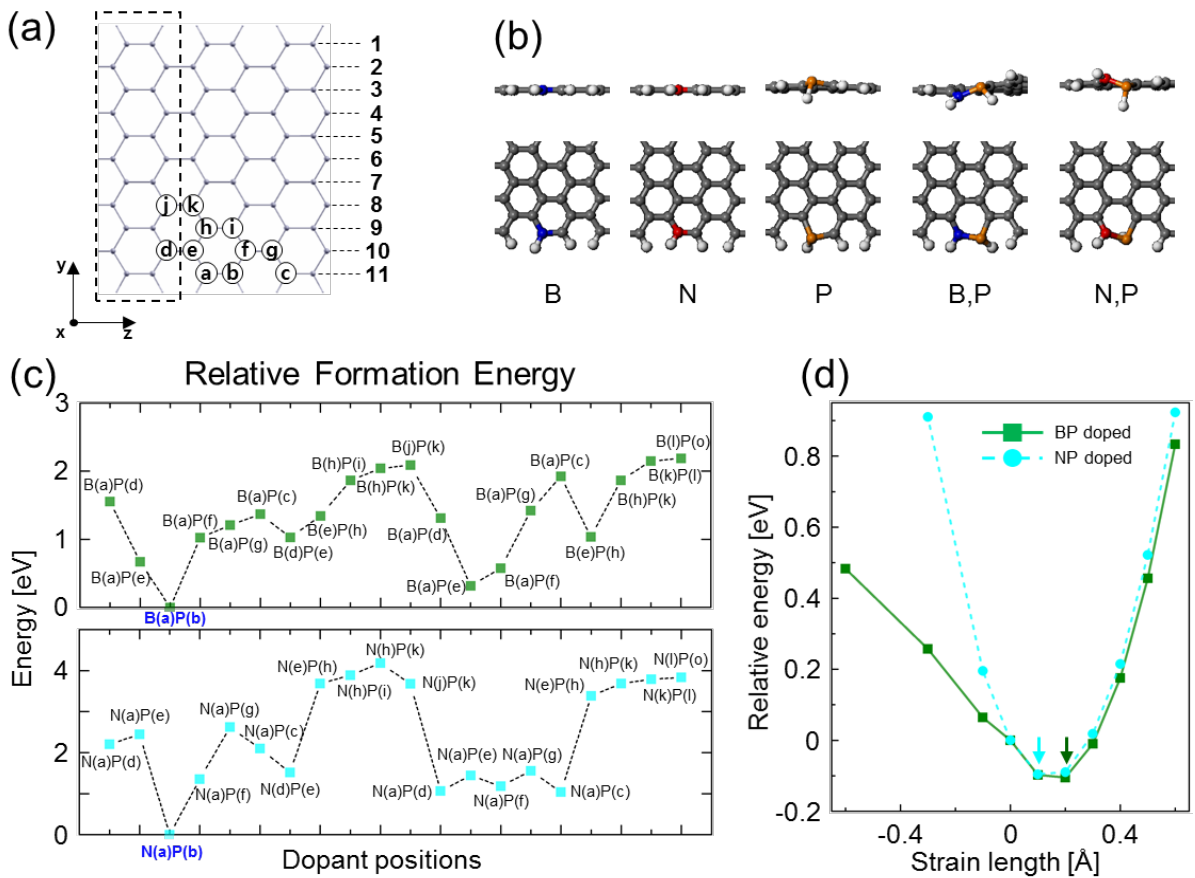

**Figure 1.** (a) Atomic structure of pristine 11-aGNRs. The dashed box indicates the unit cell with the *z*-direction cell length of 4.26 Å. Several considered doing sites are labeled as ⓐ~ⓚ. See Figure S2 for the complete catalogue of considered doping sites. (b) Side (top panels) and top views (bottom panels) of the fully optimized atomic structures of single B-, N-, P-, and binary B,P-and N,P-doped 11-aGNRs in their ground state configurations. Note the strong structural distortions occurring in the single P and binary B,P and N,P doping cases. (c) Relative formation energies of different doping positions in the B,P and N,P co-doped 11-aGNRs. See Figure S2 for the data of all considered doping cases. (d) Changes of the *z*-direction lattice constant of six unit-cell 11-aGNRs (25.56 Å) due to the B,P and N,P co-doping.

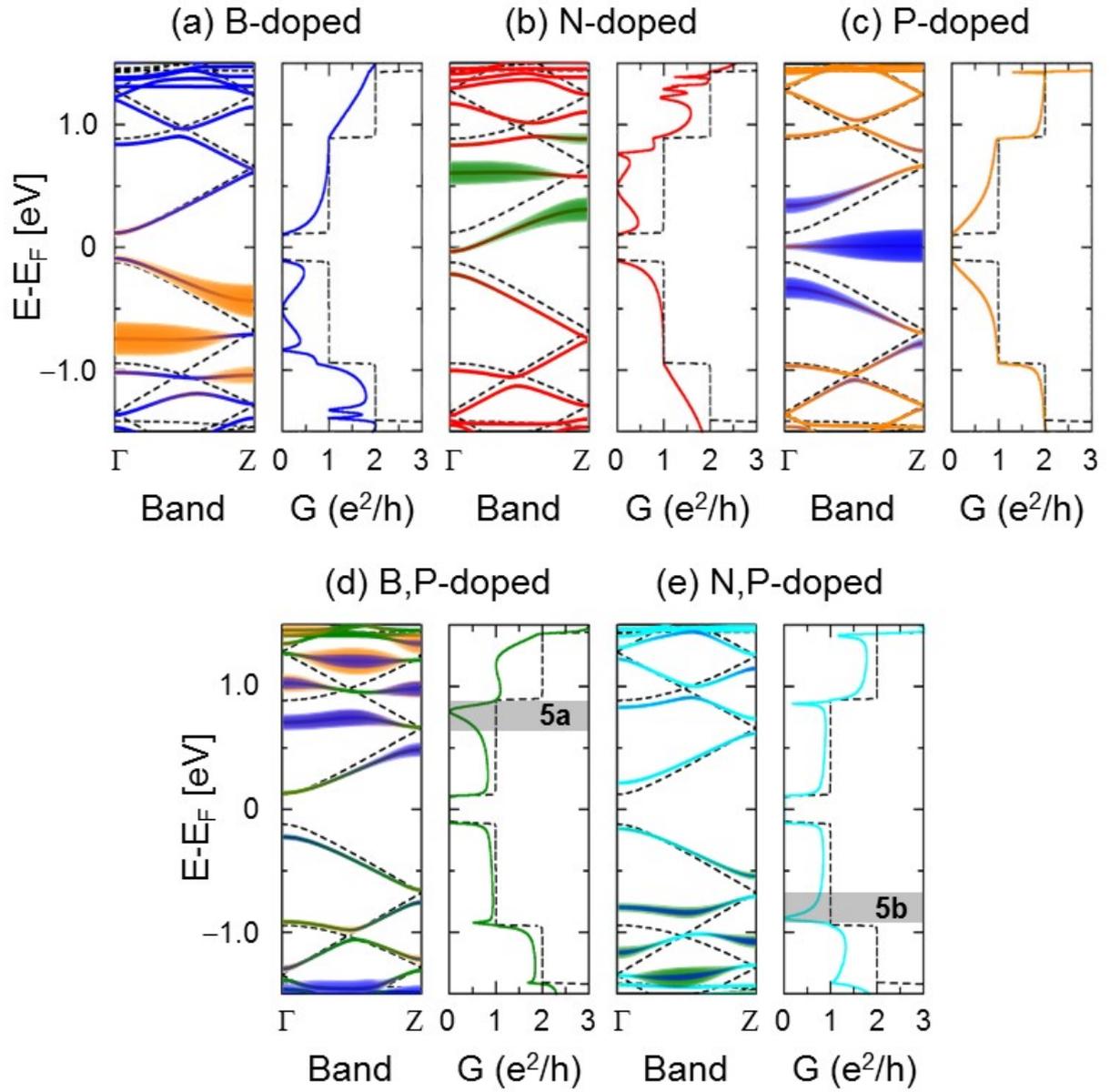

**Figure 2.** Band structure and transmissions of (a) B-doped, (b) N-doped, (c) P-doped, (d) B,P-complex-doped, and (c) N-P-complex-doped 11-aGNRs. Band structure and transmission of the pristine 11-aGNRs are shown together (dashed lines). In band structure plots, orange [in (a) and (d)], green [in (b) and (e)], and blue [in (c), (d), and (e)] shaded lines represent the B, N and P states, respectively, with the line weight corresponds to the strength of their contribution. Grey area in (d) and (e) represent the energy ranges where LDOS are plotted in Figures 5a and 5b, respectively.

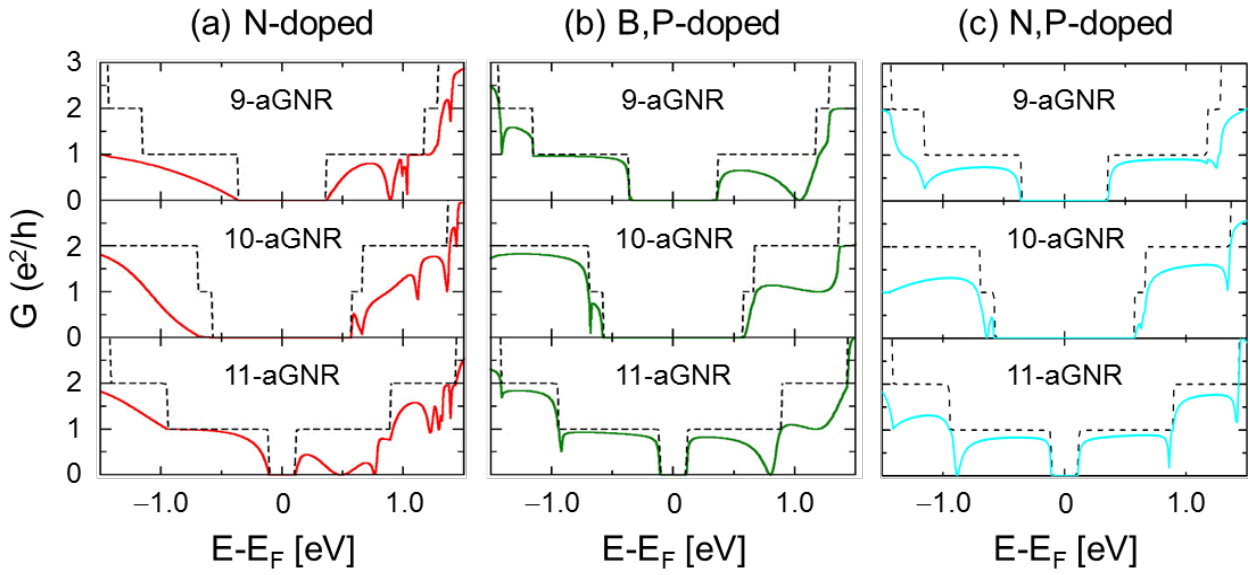

**Figure 3.** Transmissions of (a) N-doped, (b) B,P-doped, and (c) N,P-doped *N*-aGNRs (*N* = 9, 10, 11). Transmissions of pristine *N*-aGNRs are shown together (dashed lines).

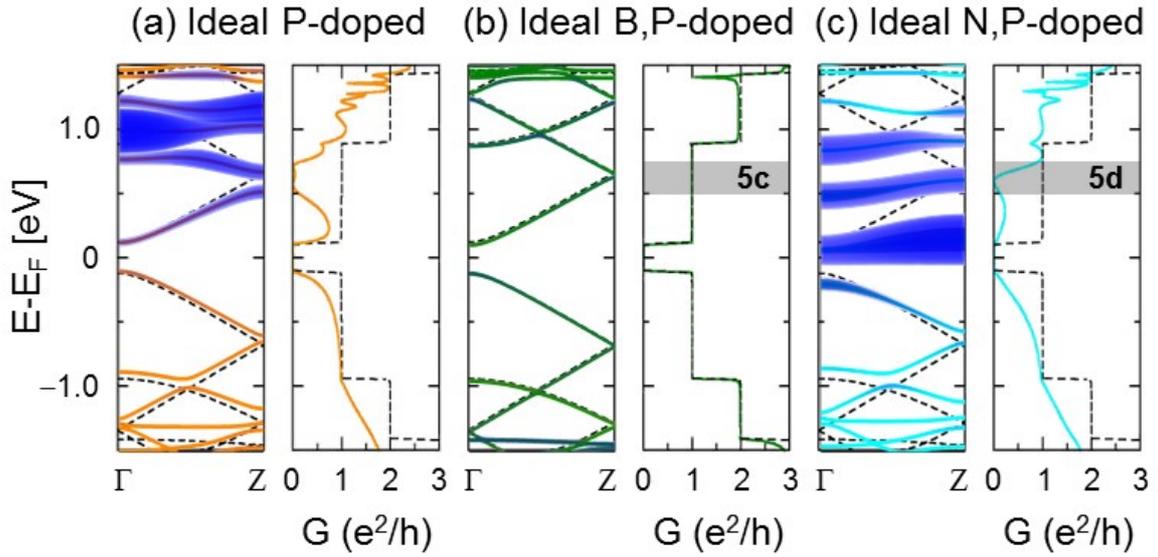

**Figure 4.** Band structure and transmissions of ideal (unrelaxed) (a) P-doped, (b) B,P-doped, and (c) N,P- doped 11-aGNRs. Band structure and transmissions of pristine 11-aGNRs are shown together (dashed lines). In band structure plots, orange, green and blue shaded lines represent the projection of B, N and P defect states, respectively, with the line weights correspond to the strength of their contribution. Grey area in (b) and (c) represent the energy ranges where LDOS are plotted in Figures 5c and 5d, respectively.

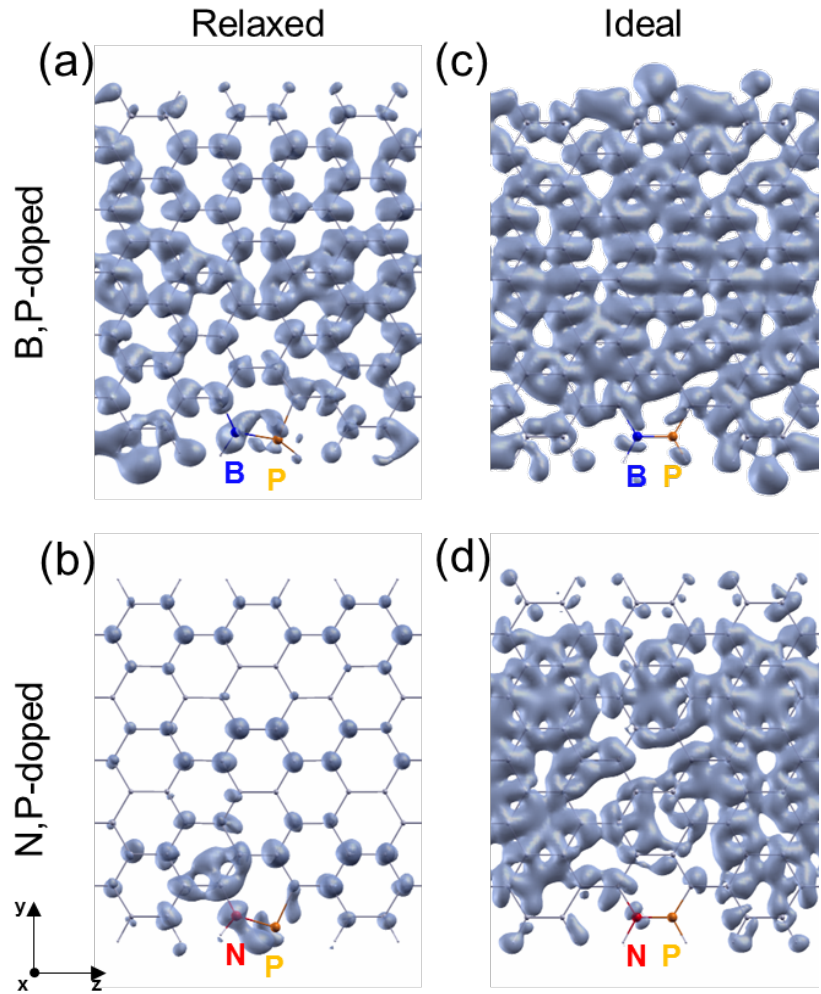

**Figure 5.** The LDOS of fully relaxed (a) B,P-complex doped and (c) N,P-complex doped 11-aGNRs for the energy ranges $E_F + 0.625$ eV ~ $E_F + 0.875$ eV (see Figure 2d) and $E_F - 0.875$ eV ~ $E_F - 0.625$ eV (see Figure 2e), respectively. The LDOS of the corresponding ideal (c) B,P-complex doped (see Figure 4b) and (d) N,P-complex doped (see Figure 4c) 11-aGNRs for the energy ranges $E_F + 0.5$ eV ~ $E_F + 0.725$ eV. The LDOS shown in (c) closely resemble the LDOS of pristine aGNR.

**Table 1.** Formation energies of the single B-, N-, and P-doped 11-aGNRs, and B,P- and N,P-co-doped 11-aGNRs.

| Single doping cases | Formation energy [eV] | Co-doping cases | Formation energy [eV] |
|---|---|---|---|
| B doping | 1.073 | B,P co-doping | 1.512 |
| N doping | − 0.749 | N,P co-doping | − 0.249 |
| P doping | 2.970 | | |